\documentclass[a4paper,12pt]{article}
\usepackage{epsfig}
\def\Journal#1#2#3#4{{#1} {\bf #2}, #3 (#4)}

\def\NPB{{\em Nucl. Phys.} B}
\def\PLB{{\em Phys. Lett.}  B}
\def\PRL{\em Phys. Rev. Lett.}
\def\PRD{{\em Phys. Rev.} D}
\def\ZPC{{\em Z. Phys.} C}

\def\be{\begin{equation}}
\def\ee{\end{equation}}
\def\bea{\begin{eqnarray}}
\def\eea{\end{eqnarray}}

\begin{document}
\begin{titlepage}
\begin{center}
\vspace*{4cm}

{\bf BOSE-EINSTEIN EFFECT FROM ASYMMETRIC SOURCES

IN MONTE CARLO GENERATORS }

\vspace{2cm}

{K. FIA{\L}KOWSKI and R. WIT}
\\

\vspace{1cm}
{\sl M. Smoluchowski Institute of Physics, 

 Jagellonian University,

ul. Reymonta 4, 30-059 Krak\'ow, Poland}

\vspace{1cm}


\abstract{
We discuss the implementations of the Bose-Einstein effect from asymmetric
sources in  Monte Carlo generators. A comparison of  LEP  data with results from the 
PYTHIA/JETSET code with the standard procedure imitating the effect and with the 
results from
the weight method
 (with weights depending in various ways on components of
momenta differences) is presented. We show that in this last method one can 
reproduce the experimental hierarchy of the source radii.}
\end{center}
\end{titlepage}

\section{Introductory remarks}
Recently one observes a renewal of interest in analysing
the space-time structure of sources in multiparticle production by means
of Bose-Einstein (BE) interference ~\cite{GGLP}. Such analysis followed 
the example of
 astrophysical investigations of Hanbury-Brown and 
Twiss~\cite{HBT}. 
The main motivation of this renewal was the analysis of the   $e^+e^-
\rightarrow W^+W^-$ process which became available at the LEP2. It was suggested 
~\cite{EG},\cite{LS} that the BE interference (and/or colour reconnection effects)
between the strings from two $W$ decays may shift the $W$ mass value fitted from 
the two jet mass distributions by as much as a few hundred MeV, thus making 
this channel useless for precise tests of the standard model. However, other
 investigations 
suggested that such a big shift is unlikely ~\cite{JZ},\cite{KKM},\cite{FW98}. Experimentally, 
the existence of interference effects between strings is still debatable  
~\cite{STN}.
\par
Investigating such subtle effects became possible when instead of the standard 
approach~\cite{BGJ} one started to model
this effect in Monte Carlo generators. There are several methods of modelling:
 as the "afterburner" for which the
original MC provides a source~\cite{SUL},\cite{zha}, by shifting the
momenta~\cite{SJO} or by adding weights to generated events~\cite{BK},\cite{FWW}.
Another approach was set forward by Andersson and collaborators who used 
the symmetrization inside fragmenting string~\cite{AH} to model the effect 
for a single string~\cite{AR}. Here we consider the most widely used methods 
of shifting momenta and weighting events.
\par
Another reason to analyze the BE effect were the efforts to estimate size and 
shape of source of particle production in various processes (in particular for 
coming RHIC data). The analysis of BE effect in 3 dimensions is supposed 
to reflect the
spatial source asymmetry.  Such analysis  was done for the LEP data at the
$Z^0$ peak~\cite{CU} which have very high statistics and good accuracy.
\par 
In this paper we compare the 3-dimensional data for BE effect from LEP  with
the results of the standard momentum shifting procedure and of the weight method.
In the next section we present the data discussing in detail the definitions and
the procedures used by the experimental groups. In the third section we compare 
them with the results obtained from the PYTHIA/JETSET MC generator using the original 
procedure modelling the effect by momentum shifting and with the results from
the weight method with
weights independent on spatial orientation of momenta. Fourth section contains the 
results for asymmetric weights. Our conclusions are presented in the last section. 

\section{Experimental data}
Although the discussion of the shape of asymmetric sources in the framework
of BE interference concerned most often the heavy ion collisions, the best 
experimental data with highest statistics exist for the $e^+e^-$ annihilation 
at the $Z^0$ peak. 
In the following we concentrate our attention
on the L3 data~\cite{L3} which discuss the ratios using "uncorrelated background"
and three different radii to parametrize the data. The DELPHI data~\cite{DEL} 
are parametrized
with only two radii, and the OPAL data~\cite{OPAL} use the like/unlike
ratio which requires a cut off of the resonance affected regions even in
double ratios.
\par
As in the L3 paper~\cite{L3} we use for each pair of identical pions
three components of the invariant
$Q^2={-(p_1-p_2)^2}$: $Q_L^2, Q_{out}^2, Q_{side}^2$ defined in the LCMS 
(Longitudinal Center-of-Mass System), 
where the sum of three - vector momenta is
perpendicular to the thrust axis. The $Q_{out}$ component is measured along this sum,
the $Q_L$ along the thrust axis, and $Q_{side}$ is the projection of $Q$ on the axis
perpendicular to these two directions ~\cite{CP},\cite{L3}.
 
We  define a "double ratio"
 in the same way as in the L3 paper 
using a  reference sample from mixed events:  

\begin{equation}
R_2(p_1, p_2) =
\frac{\rho_2}{\rho_2^{mix}} / \frac{\rho_2^{MC}}{\rho^{mix, MC}_{2}}.
\label{eq:double}
\end{equation}

 This "double ratio" is
  parametrized by 

$$ 
R_2(Q_L,Q_{out},Q_{side}) = \gamma [1+ \delta Q_L+
 \epsilon Q_{out}+ \zeta Q_{side}]~\cdot
 $$ 
\begin{equation}
\cdot ~ [1+ \lambda \exp(-R^2_L Q_L^2-R^2_{out}Q^2_{out}-
R^2_{side}Q^2_{side}-
 2\rho_{L,out}R_{L}R_{out} Q_{L} Q_{out})]
.
\label{eq:fit}
\end{equation}

The first bracket reflects possible traces of
long-distance correlations; the last term in the second bracket seems to
be negligible when fitting data and will be omitted in the following.
 
\par
By fitting the parameters $R_L$ and $R_{side}$ we get some information on
the geometric radii in the longitudinal and transverse directions
(respective to the thrust axis). $R_{out}$ reflects both the spatial
extension and time duration of the emission process.
\par
In the L3 data the fit region in all three variables extends to
$1.04~ GeV$ and is divided into 13 bins, which gives 2197 points fitted with
8 parameters.  The fit parameters $\delta, \epsilon$ and $\zeta$ are
rather small; this means that the observed BE enhancement is rather well
approximated with a Gaussian.  The 
value of the parameter $\lambda$ is fitted as $0.41 \pm 0.01$. 
\par The fitted values of radii (in  $fm$) are as follows: 
 $$ R_L = 0.74 \pm 0.02^{+0.04}_{-0.03},~ R_{out} =
0.53 \pm 0.02^{+0.05}_{-0.06},~ R_{side} = 0.59 \pm 0.01^{+0.03}_{-0.13}$$
We see clear evidence for source elongation:
$R_{side}/R_L$ is smaller than one by more than four standard deviations.
\par
\begin{figure}[h]
\centerline
{
\epsfig{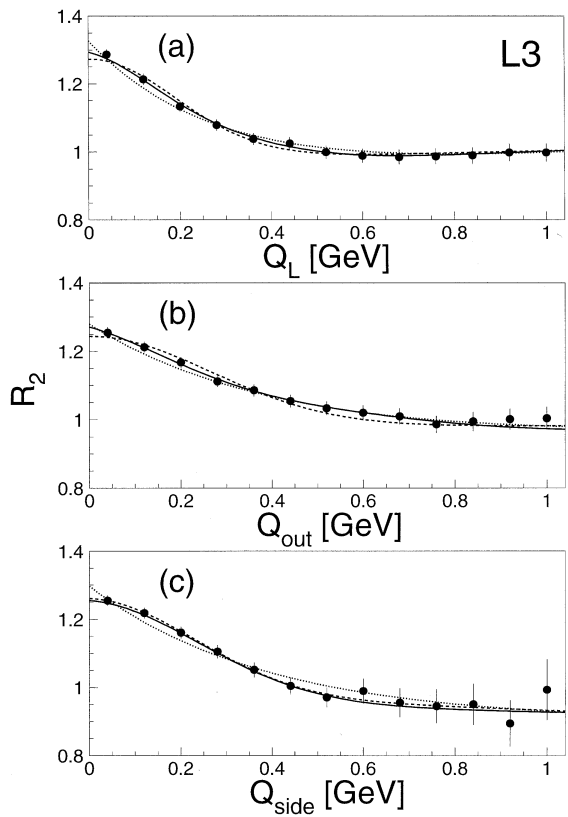}}
\caption{Projections of the double ratio (\ref{eq:double}) from the data of
the L3 collaboration on the three axes $Q_L$, $Q_{out}$ and $Q_{side}$.}
\label{fig:fig1}
\end{figure}

It is instructive to inspect the projections of the double ratio on the
three axes $Q_L$, $Q_{out}$ and $Q_{side}$. This is done by restricting 
the values of two other variables to less than $0.24~GeV$, plotting the
histograms in the third variable in bins of width $0.08~GeV$ and constructing 
the double ratio in this variable. The results are shown in Fig. \ref{fig:fig1}
 as presented by the L3 collaboration \cite{L3}.

The values of double ratios fall down smoothly from the maxima of about 1.25
at $Q$-s close to zero to the plateau at 1. It is rather difficult to see
the differences between three plots, but superposing them one may note 
that the fall is fastest for $Q_L$ as expected from the fact that the fitted
value of parameter $R_L$ is bigger than the values of $R_{out}$ and $R_{side}$
quoted above. 

\section{Asymmetric effects from symmetric models}
The geometric interpretation of data requires  a comparison with the
results from the
standard MC procedures modelling the BE effect.  In the L3 paper such
an analysis is given for the standard LUBOEI procedure built into the
JETSET Monte Carlo generator. This procedure modifies the final state by a
shift of momenta for each pair of identical pions. The shift is calculated
to enhance low values of $Q^2$ and to reproduce the experimental ratio in
this variable. The function defining this shift is 
\begin{equation}
f(Q^2) = 1 + \lambda_{in} \exp(- R_{in}^2Q^2)
\label{eq:gaus}
\end{equation}

 The superposition of the procedure for all the pairs and
subsequent rescaling (restoring the energy conservation) makes  the
connection between the parameters of the shift $\lambda_{in}, \ R_{in}$ 
and the parameters describing the  resulting double ratio
 in $Q^2$ 

\begin{equation}
R_2(Q^2) =
 \frac{\rho_2}{\rho_2^{mix}} / \frac{\rho_2^{MC}}{\rho^{mix, MC}_{2}}
\label{eq:gaus2}
\end{equation}
(which may be parametrized analogously to (\ref{eq:gaus})) rather indirect. 
 
\par 
Using the JETSET parameters adjusted to all the L3
data and the LUBOEI parameters fitted to describe the BE ratio in $Q^2$
the authors of the L3 paper calculated the same quantities as measured in
the experiment.  The projections of $R_2$ are qualitatively very similar
to the experimental ones. However, the fit to the 3-dimensional
distribution gives results different from data. The ratio $R_{side}/R_L$
is not smaller but greater than one; the fitted values (in $fm$) are:
$$R_L= 0.71 \pm 0.01, R_{out} = 0.58 \pm 0.01, R_{side} = 0.75 \pm 0.01.$$
We confirmed these numbers in our calculations. We found also
that the results are  sensitive to the JETSET parameters. Using the
default values instead of the L3 values we obtained
a significantly smaller value of $R_{out}$ (below 0.5) and significantly
smaller $\lambda$. Other values are less affected and $R_{side}/R_l$ is still
bigger than 1.
\par
\begin{figure}[h]
\centerline
{
\epsfig{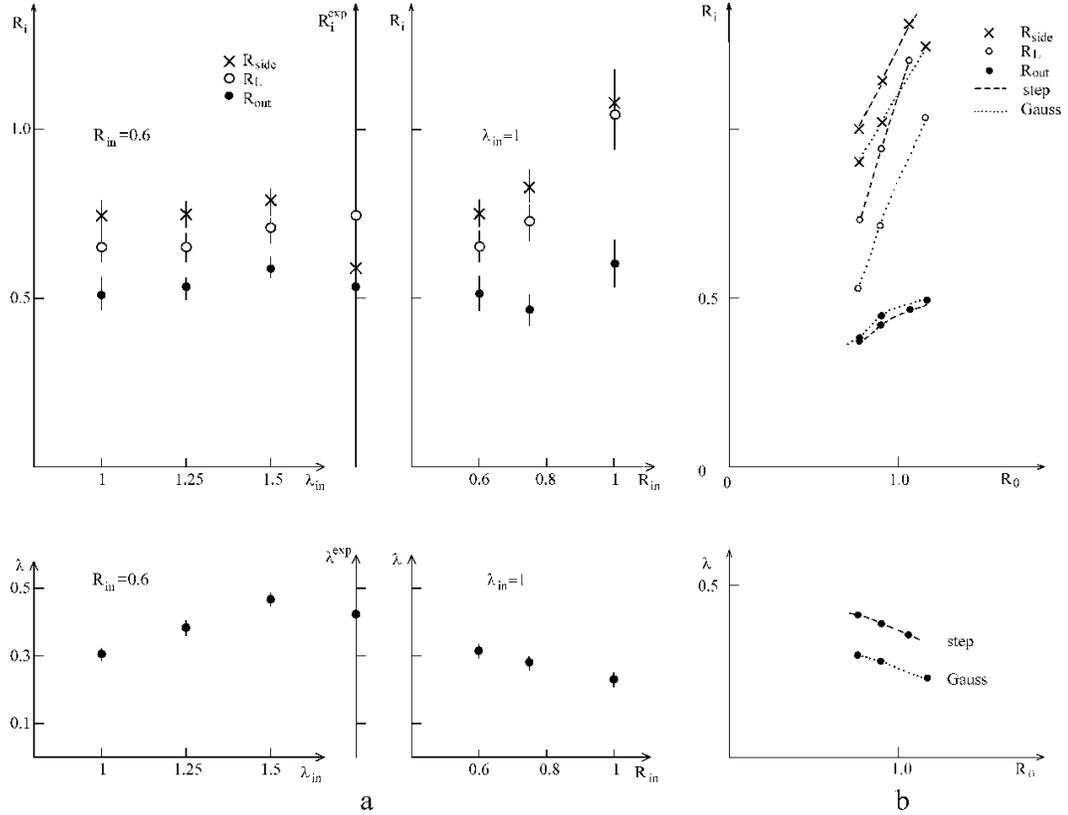}}
\caption{Fit parameters $\lambda$ and $R_i$ as functions of the
input parameters a) for the LUBOEI procedure, b) for the weight method.
Experimental values are shown on a separate vertical axis.}
\label{fig:fig2}
\end{figure}

We have also checked  how the results depend on the source radius $R_{in}$ and
on the incoherence parameter $\lambda_{in}$ assumed in the LUBOEI input 
function
 (\ref{eq:gaus}).  In all cases we get
$R_{side}>R_L>R_{out}$, although the input function was obviously
symmetric.  The values of $R_{side}$ and $R_L$ are proportional to $R_{in}$,
whereas $R_{out}$ changes much less; the dependence on $\lambda_{in}$ is
very weak.  The output value of $\lambda$ decreases quite strongly with
increasing $R_{in}$ and increases with $\lambda_{in}$. No choice of input
parameters gives the values of $R_i$ compatible with data.  This is shown
in Fig.~\ref{fig:fig2}.

\par Another interesting observation is that to
fit the L3 data one needs $\lambda = 1.5$, which is beyond the physically
acceptable value of 1. This supports our doubts about usefulness of
the LUBOEI procedure in understanding the experimental results (although
certainly it is the most practical description of data).
 
\par
In fact, there is one more degree of freedom in the prescription for 
modelling the BE effect: the definition of direct pions. Since the decay 
products of long-living resonances and of particles decaying by 
electroweak interactions are born far from the original collision point, 
their effective source size is much bigger than that for direct pions. Thus 
they contribute to the BE effect for momentum differences much below 
the experimental resolution and should not be taken into account.
\par
In the LUBOEI procedure this distinction is made by the decay width 
of unstable particles: only pions from the decay of particles with the 
width above $20MeV$ and the direct ones are included in the momentum shifting
procedure. Obviously, this is just a rough prescription which may be changed,
and the values of fit parameters may change then quite strongly. The user of 
the procedure should be aware that (according to author's warning) it works 
properly only when called from LUEXEC; if LUBOEI is called directly from the
master program, all pions are regarded as the direct ones. 
\par 
The problems of LUBOEI procedure in describing the asymmetry of experimental
distributions are not the first ones noted in applications to describe various data.
It has been already indicated that the procedure with parameters fitted to the
two-particle data fails to reproduce the three-particle spectra~\cite{Wu} and the 
semi-inclusive data~\cite{NA22}. Moreover, as already noted, the fitted values of 
parameters needed in the input function~(\ref{eq:gaus}) are quite different from 
the values one would get fitting the resulting double ratio~(\ref{eq:gaus2}) to the 
same form~\cite{FW97},\cite{SLM}. Thus, it seems to be  difficult to learn something 
reliable on the space-time structure of the source from the values of the fit
parameters in this procedure. 
\par
All this led to a revival of weight methods, known for quite a long time~\cite{Pratt},
but plagued also with many practical problems. The method is clearly justified 
with
in the formalism of the Wigner functions, which allows one to represent
(after some simplifying assumptions) any distribution with the BE effect
built in as a product of the original distribution and the weight factor,
depending on the final state momenta~\cite{BK}. With an extra assumption on 
factorization in momentum space we may write the weight factor for a final
state with $n$ identical bosons as 

\begin{equation} W(p_1,...p_n)=\sum
\prod_{i=1}^n w_2(p_i,p_{P(i)}),
\label{eq:weight}
\end{equation}

\noindent
where the sum extends over all permutations ${P_n(i)}$ of $n$ elements, and
$w_2(p_i,p_k)$ is a two-particle weight factor reflecting the effective source
size. Problems with an enormous number of possible terms in this sum may be cured by
a proper clustering procedure~\cite{FWW}. A reasonable description of the effect
in $Q^2$ is obtained with a simple gaussian form of the weight factor
\begin{equation}
w_2(p_1,p_2)= \exp[-(p_1-p_2)^2R_{in}^2/2],
\label{eq:wf}
\end{equation}
or, even simpler, a step function form with $w_2 = 1 $ for some
range of $-(p_1-p_2)^2<1/R_{in}^2$ and $w_2=0$ outside \cite{FW00}.
\par
In this method we may repeat the same calculation as done for the LUBOEI
procedure. Obviously the weights may be calculated for the events generated 
by any MC generator, but here we restrict ourselves to the results from the 
same PYTHIA/JETSET code which was used above.
 The resulting double ratios are not that smooth and monotically
decreasing as in the data or from the LUBOEI procedure (which is the
usual drawback of the weight methods).  However, the major features are
surprisingly similar:  with weight factors depending only on $Q^2$ we get
different values of fitted $R_i$ parameters.  Moreover, the hierarchy of
parameters is the same:  $R_{side} > R_L > R_{out}$. This suggests that the
assymetry is generated by the jet-like structure of final states
and not by any specific features of the procedure modelling the BE effect.
In Fig. 2b we show the values of the fit parameters as functions of $R_{in}$
for a Gaussian as well as  the $\theta$-like weight factors.
Again, no choice of the input parameters allows to describe the data.
\par
The comparison of two methods is not straightforward. In particular, one
should take care if the same definition of "direct" pions is used.  
The weights are calculated after the event was fully generated (and all the
decays of unstable particles occured). Therefore one should define the pions 
which are counted as direct ones. We did it by enumerating particles, which
contribute significantly to the pion production and live too long 
for their decay products to produce a visible  BE effect (using 
the same limit for decay width as in LUBOEI). If one
enumerates the short-living resonances and adds their decay products to the
direct pions, one should remember that this list is different in various 
options of JETSET (e.g. the option used by the L3 collaboration 
takes into account mesons built from
quarks with non-zero orbital momentum, which are neglected in the 
default version).
\par
The results presented in this section suggest that one should be careful
with the geometric interpretation of the data. If one gets asymmetric
distributions from the generator without assuming explicitly space
asymmetry of the source, it is not clear how the assumed asymmetry will be
reflected in the results.

\section{Asymmetric weights}
One may get more information on the problem of asymmetric BE effect in MC 
generators using the asymmetric weight method, i.e.  introducing weight
factors which depend in a different way on $Q_L = |p_{1L} - p_{2L}| $, 
$Q_{side}= |p_{1side} - p_{2side}|  $ and
$Q_{out}= |p_{1out} - p_{2out}|$, where the indices denote the components 
defined in the previous section. We have used two such generalizations 
of a gaussian weight factor (\ref{eq:wf}) 

\begin{equation}
w_2(Q_L, Q_{out}, Q_{side}) = \exp([-Q_L^2(R^{in}_L)^2 -Q_{out}^2(R^{in}_{out})^2 -
Q_{side}^2(R^{in}_{side})^2]/2)
\label{eq:asym1}
\end{equation}

and

\begin{equation}
 w_2(Q_L, Q_{out}, Q_{side}) = \exp([-Q_L^2(R^{in}_L)^2 -(1-\beta ^2)
    Q_{out}^2(R^{in}_{out})^2 - Q_{side}^2(R^{in}_{side})^2]/2)
\label{eq:asym2}
\end{equation}

where $\beta$ is defined as

\begin{equation}
      \beta = \frac{p_{out1} + p_{out2}}{E_1 + E_2}.
\end{equation}

\par
The weight factor (\ref{eq:asym2}) reduces to the symmetric weight factor 
(\ref{eq:wf}) when $R^{in}_L=R^{in}_{out}=R^{in}_{side}=R_{in}$.
The formula  (\ref{eq:asym2}) gives nearly the same results as the formula 
(\ref{eq:asym1}) when $R_{out}$ is multiplied by 2. We have used both forms
finding no definite preference for any of them. 
\par
Fluctuations in the 
weight values are large and the resulting fluctuations in the values 
of double ratios describing the BE effect are bigger than for the momentum 
shifting method. Therefore it is necessary to use large samples of generated 
events. We found that for the samples of 5 million events the fluctuations 
visible in the plots of projections of double ratios on components of $Q$ 
are comparable with those seen in the experimental data shown in Fig.1. 
In fact, the plots obtained for the weight method with the input radii around 
$0.5 ~fm$ are visually similar 
to those of experimental data. However, the fitted values of the parameters 
from formula (\ref{eq:fit}) are different. 
\par 
Since for the symmetric weights the resulting fitted values of 
$R_{side}$ are bigger than the values of $R_L$ (contrary to the inequality 
seen in the data), it seemed natural to take the input value of $R^{in}_{side}$
smaller than $R^{in}_L$. Indeed decreasing $R^{in}_{side}$ one reduces the resulting
fitted value of $R_{side
}$, but this dependence is not linear and saturates for
$R^{in}_{side}$ around $0.3~fm$. Moreover, the fitted values of other parameters
change as well although their input values were not changed. Therefore  looking
for the best set of input parameters in the formula for weights is a rather involved
procedure.
\par Let us add two more remarks. A replacement of  
the products of Gaussians  by the
proper products of step functions in the formulae for weights (\ref{eq:asym1}), (\ref{eq:asym2})
leads to even bigger fluctuations in the 
resulting distributions and we do not advocate such parametrizations. Finally, there
is some ambiguity concerning the use of weights for the calculations of double ratio 
(\ref{eq:double}). If we use the weights only for the two-particle distributions, the
two denominators cancel and we calculate effectively just the ratio of two-particle
distributions with- and without weights. It seems, however, that the justification
for the weight method ~\cite{BK} requires using weights both for the single- and 
two-particle distributions. We have looked for the best set of parameters with this
prescription, using a Gaussian form without the "$\beta$-factor" (\ref{eq:asym1}).
The best set we found is 

\begin{equation}
R_L^{in} = 0.9 fm, R_{out}^{in} = 0.3 fm, R_{side}^{in} = 0.4 fm.
\label{eq:param}
\end{equation}  

The resulting projections of the double ratios are shown in Fig.3. The fitted values
of parameters in formula (\ref{eq:fit}) we get are

\begin{equation}
R_L = 0.73 fm, R_{out} = 0.54 fm, R_{side} = 0.65 fm.
 \label{eq:param2}
\end{equation}  

Obviously, it is now possible to reproduce the experimental hierarchy of radii. The fitted
value of $\lambda$ is smaller than in data ($0.35$ instead of $0.41$), but the difference
is well within the systematic errors of the fit to the experimental data. Note that we are not
showing the errors in Fig.3 (nor quoting them in the values of parameters listed above), 
since these errors result mainly from the fluctuations in weights. Some estimate
 is
obtained by comparing the results for 1 and 5 million events samples; in Fig.3 the
differences are of the order of size of the points. 

\begin{figure}[h]
\centerline{
\epsfig{file  = 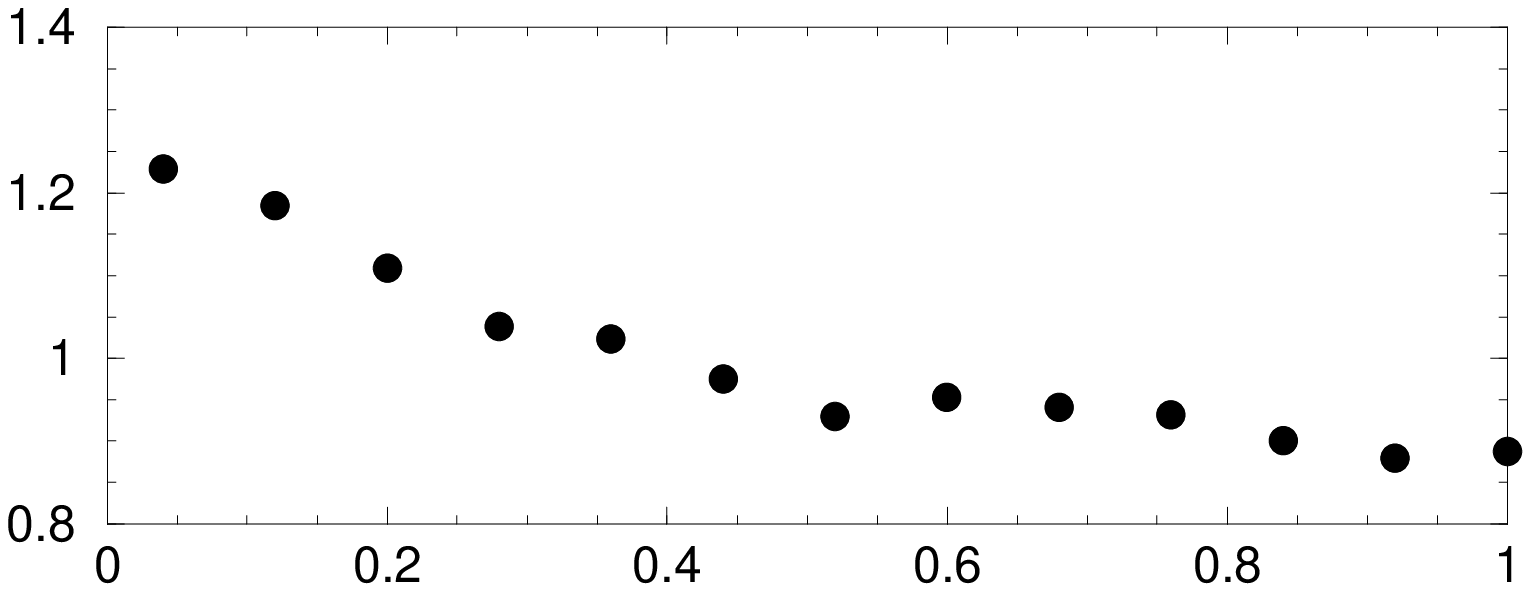,height=4cm}
}
\label{fig:fig3a}
\end{figure}
\centerline{\bf $Q_L$}
\begin{figure}[h]
\centerline
{
\epsfig{file= 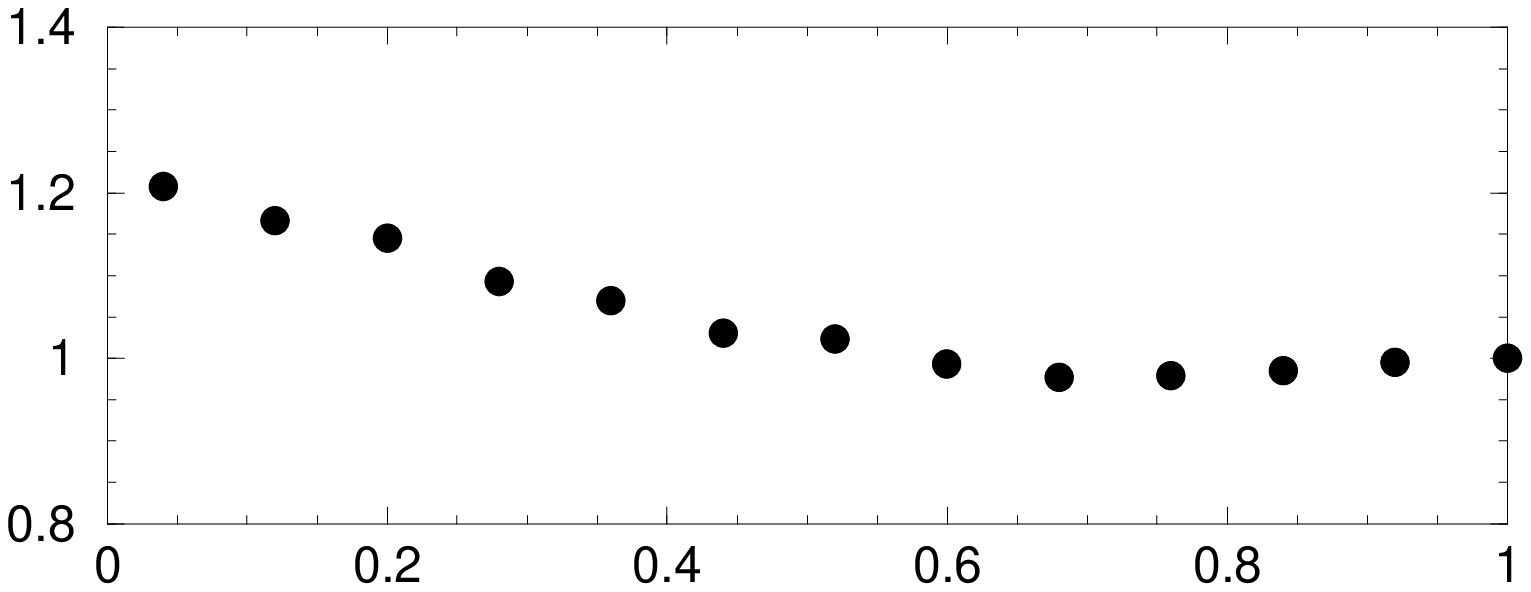,height=4cm}
}
\label{fig:fig3b}
\end{figure}

\centerline{\bf $Q_{out}$}
\begin{figure}[h]
\centerline{
\epsfig{file = 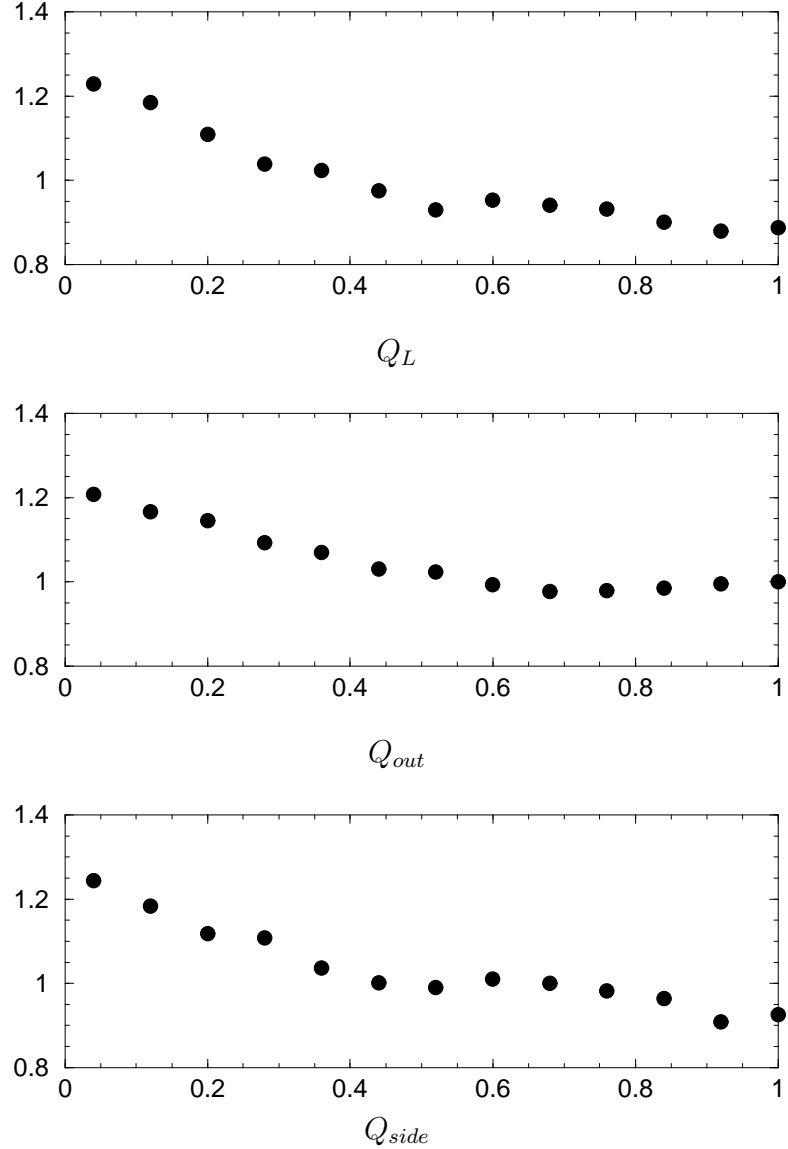,height=4cm}
}
\centerline{\bf $Q_{side}$}
\caption{Projections of the double ratio (\ref{eq:double}) from 
the PYTHIA/JETSET MC generator with the asymmetric weight 
method for parameters (\ref{eq:param2}) on the three 
axes $Q_L$, $Q_{out}$ and $Q_{side}$.}
\label{fig:fig3c}
\end{figure}
\par There is a striking difference between the input values of the radii 
(\ref{eq:param}) assumed in the weight factors and the resulting best fit
values (\ref{eq:param2}) from the double ratio calculated with these weights. 
Although the hierarchy $R_L > R_{side} > R_{out}$ is the same in both cases, 
the fitted values differ by less than 25\%, whereas there is a difference by 
more than a factor of two between the input values. 
\par 
Moreover, further decrease of  the values of $R_{out}^{in}$ and $R_{side}^{in}$
 hardly affects the resulting double ratio and fitted values 
of $R_i$. This seems to be the inherent property of the JETSET generator,
 which yields a rather strong suppression of large values of $ Q_i$ and 
 $Q^2$ even without any procedure imitating the BE effect. 
 Apparently this suppression dominates over the weak enhancement of low values
 of $Q_i$ induced by the weight factors with small values of $R_i$. For  
small $R_i^{in}$ there is no simple correspondence between the input and output 
values of radii.  This looks analoguos to the effect noted already for a symmetric 
BE effect described  by the LUBOEI procedure~\cite{SLM}. Therefore any direct 
interpretation of the fit values for BE double ratios in terms of the different 
radii of the asymmetric source is a rather delicate matter.

\section{Conclusions}
In this note we present the results of our investigation 
concerning  the asymmetry of the BE effect in two
procedures imitating this effect in the Monte Carlo generators. A
comparison with the data  at $Z^0$ peak is presented. We found that 
both the momentum shifting method and the weight method with weights 
depending on $Q^2$  only give different distributions in different 
components of $Q^2$. However, the hierarchy of radii parametrizing 
these distributions is different from the experimental one. Introducing 
weights which depend in different ways on different components 
of $Q^2$ we are able to reproduce the experimental data.
 
\section*{Acknowledgments}
Thanks are due to A. Bia{\l}as for reading the manuscript. This work was 
partially supported by the KBN grants No 2 P03B 086 14, 
2
 P03B 010 15 and  2 P03B 019 17.

\end{document}